\documentclass[aps,prev,twocolumn,preprintnumbers,floatfix,nofootinbib]{revtex4-1}
\pdfoutput=1
\usepackage{graphicx}
\usepackage{bm}
\usepackage{times}
\usepackage{slashed}
\usepackage{color}
\usepackage{slashed}
\usepackage{lipsum}
\usepackage{subfigure}
\newcommand{\be}{\begin{equation}}
\newcommand{\ee}{\end{equation}}
\newcommand{\bea}{\begin{eqnarray}}
\newcommand{\eea}{\end{eqnarray}}

\begin{document}

\title{Dark Matter and Global Symmetries}

\author{Yann Mambrini$^1$}
\email{yann.mambrini@th.u-psud.fr}

\author{Stefano Profumo$^2$}
\email{profumo@ucsc.edu}

\author{Farinaldo S. Queiroz$^{2,3}$}
\email{Farinaldo.Queiroz@mpi-hd.mpg.de}

\affiliation{$^1$Laboratoire de Physique Theorique, Universit\'e Paris-Sud, F-91405 Orsay, France\\
$^2$Department of Physics and Santa Cruz Institute for Particle
Physics University of California, Santa Cruz, CA 95064, USA\\
$^3$Max-Planck-Institut f\"ur Kernphysik, Postfach 103980, 69029 Heidelberg, Germany}

\begin{abstract}
 
General considerations in general relativity and quantum mechanics are known to potentially rule out continuous global symmetries in the context of any consistent theory of quantum gravity. Assuming the validity of such considerations, we derive stringent bounds from gamma-ray, X-ray, cosmic-ray, neutrino, and CMB data on models that invoke global symmetries to stabilize the dark matter particle. We compute up-to-date, robust model-independent limits on the dark matter lifetime for a variety of Planck-scale suppressed dimension-five effective operators. We then specialize our analysis and apply our bounds to specific models including the Two-Higgs-Doublet, Left-Right, Singlet Fermionic, Zee-Babu, 3-3-1 and Radiative See-Saw models. {Assuming that (i) global symmetries are broken at the Planck scale, that (ii) the non-renormalizable operators mediating dark matter decay have $O(1)$ couplings, that (iii) the dark matter is a singlet field, and that (iv) the dark matter density distribution is well described by a NFW profile}, we are able to rule out fermionic, vector, and scalar dark matter candidates across a broad mass range (keV-TeV), including the WIMP regime.

\end{abstract}


\maketitle

\section{Introduction}
Particle physics models achieve stability for dark matter (DM) particle candidates by advocating the presence of either discrete or continuous global symmetries. Discrete symmetries arise, for example, from broken gauge (local) symmetries, which are respected at the Planck scale \cite{Martin:1992mq,Krauss:1988zc}. 
Continuous global symmetries, instead, are generically violated at the Planck scale, leading to important implications on the dark 
matter phenomenology of the associated models. 

There are several reasons why continuous global symmetries are not expected to be present in a consistent theory of quantum gravity, which rely on general facts in gravity and quantum mechanics:

(i) {\it No-Hair Theorem}: Since local $U(1)$ symmetries are effectively identical to Gauss's law, any observer outside a Black Hole (BH) horizon can determine the BH charge. However, if there existed continuous global symmetries, when a charged particle gets trapped inside the BH there would be no way to assess this from outside the horizon. Thus the charge would appear to be ``deleted'', in contradiction to its conservation \cite{Kallosh:1995hi}.

(ii) {\it Hawking Radiation}: The main problem with continuous global symmetries has to do with Hawking radiation \cite{Hawking:1974sw}. Since there are no gauge 
interactions associated with continuous global symmetries, one could throw a large amount of charged particles into a BH and increase its charge ($Q$)
 indefinitely \cite{Hawking:1974sw,Banks:2006mm}. However, the theory of Hawking radiation indicates that until $T_{\rm Hawking} > m$, where $m$ is the mass of the lightest 
 charged particle pair, the BH does not radiate charge. Combining this with the bound on the BH mass, namely $Qm \leq M_{BH} \leq M_{pl}^2/m$, 
 where $M_{pl}=1.22 \times 10^{19} GeV$ is the Planck mass, we find $Q \leq M_{pl}^2/m^2$. This limit can however be violated by making $Q$ sufficiently large. 
 Hence, if $Q$ were conserved we could have identical BHs with an infinite number of states labelled by $Q \gg M_{pl}^2/m^2$.

(iii) {\it Entropy}: Since an external observer cannot infer a global charge, in order to assign an entropy to a given BH one would have to count all micro-states of all charges, finding an entropy of order $\sim\log(Q)$. Now, taking $Q$ indefinitely large, one would violate the Bekenstein-Hawking formula, which says that entropy counts the number of states of a BH. Therefore, such objects are ruled out, as are continuous global symmetries \cite{Banks:2006mm}. \\

While there are general arguments for the breaking of continuous global symmetries at the scales of quantum gravity, those are not well-established. For example, Ref.~~\cite{Dvali:2015aja} disputes such asguments based on non-thermal deviations from Hawking radiation. This notwithstanding , hereafter, we assume that continuous global symmetries are indeed broken at the Planck scale and show that the notion of such symmetries being broken at the Planck scale has profound implications on DM phenomenology. \\

In this study, we assess the possibility of using continuous global symmetries to stabilize DM particles. In order to derive results applicable to a variety of particle physics models, we consider Planck-scale suppressed, dimension-five effective operators that mediate the decay of generic DM particles of spin 0, 1/2 and 1; the operators under consideration violate continuous global symmetries, and thus induce the decay of DM particles whose stability relies on such continuous global symmetries. The decay of long-lived but metastable DM particles can inform us on the DM particle nature (see e.g. \cite{Hambye:2010zb,Boucenna:2012rc}); for example, stringent bounds on the lifetime of electroweak-scale DM stem from the observed diffuse gamma-ray flux \cite{Cirelli:2012ut}, which implies lifetime $\tau \gtrsim 10^{26}$ s, thus a billion times longer than
 the age of the Universe. We emphasize the fact that even though continuous global symmetries might break down to $Z_N$ discrete symmetries at low energies, one can always generically construct Planck-suppressed effective operators that would induce the decay of the DM particle: our results can thus be applied to any continuous global symmetry.\\

Naively, one might expect that Planck-scale suppression might have a negligible impact on the phenomenology of models which advocate the existence of continuous global symmetries to stabilize the DM particle. Using current cosmic-ray, X-ray, gamma-ray, neutrino and CMB data, spanning the entire keV-TeV energy range, we show that, somewhat surprisingly, continuous global symmetries are strongly disfavored as a mechanism to stabilize DM particles. In particular, we rule out, under the aforementioned assumptions,  a rather large DM mass range, including the classic WIMP mass range around the electroweak scale.

\section{Observational Constraints}

In this section we summarize how we derive our model-independent limits on the DM lifetime. We employ throughout our analysis an NFW profile \cite{Navarro:1996gj},
\begin{eqnarray}
\rho(r) = \frac{ \rho_s}{r/r_s (1+r/rs)^2},
\end{eqnarray} with a scale radius, $r_s=24.42$~kpc, and $\rho_s=0.184$, such that to reproduce a local density of $\rho_{\odot}=0.4\,{\rm GeV/cm^3}$ \cite{Catena:2009mf}. We point out that our results would all scale linearly with other choices for the local dark matter density, unlike for the dark matter pair-annihilation case.

I suspect the limits will essentially 
 stay the same, since DM decay signals at high latitudes are not very 
 sensitive to this.

\subsection{CMB data}

Precise measurements of the Cosmic microwave background (CMB) provide robust limits on DM decays, since the latter alter the ionization and heating history of the CMB as well as its power spectrum. Using combined data from Planck \cite{Ade:2013zuv}, WMAP9 \cite{Hinshaw:2012aka}, Atacama Telescope \cite{Sievers:2013ica}, South Pole Telescope \cite{Hou:2012xq}, Hubble Space Telescope \cite{Riess:2011yx} and Baryonic Acoustic Oscillations \cite{Anderson:2012sa}, we derive limits on the DM lifetime for several final states.  Typically, those limits come from constraints on new sources of ionization and heating stemming from the products of DM interactions. Here we pay special attention to DM particle decays. Our findings rely on several standard assumptions namely: (i) the DM lifetime is large than the age of the universe; (ii) the DM particle accounts for the DM cosmological abundance; (iii) the DM particle decays fully to SM particles; (iv) the energy fraction which the DM particle deposit into the intergalactic Medium is determined by the transfer functions provided in Ref.~\cite{Slatyer:2012yq}.  The rate at which a given DM particle decay induces heating and ionization of the baryonic component of the IGM is proportional to

\begin{equation}
\Gamma(z)= \frac{1}{H(z)(1+z)n_H(z)}\left(\frac{dE}{dt dV}\right)_{\rm d}
\end{equation} where $H(z)$ is the Hubble rate and $n_H (z) = n_{H_o} (1+z)^3$ is the number density of hydrogen in the Universe at a given redshift with $n_{H_o}=1.9 \times 10^{-7} {\rm cm^{-3}}$ being the present-day value, and

\begin{eqnarray}
&&\left(\frac{dE}{dtdV}\right)_{d} = 13.7 \times 10^{-24} \left(\frac{f_{dec}}{0.1}\right) \left(\frac{n_{H_o}}{1.9 \times 10^{-7}}\right) \left( \frac{\Omega_{DM} h^2}{0.13}\right) \nonumber\\
&&\left( \frac{10^{25} s}{ \tau_{DM}} \right) (1+z)^3 {\rm eV/s} \nonumber\\.
\end{eqnarray} where $f_{dec}$ is in general a function of the DM mass and redshift. We closely follow Ref.~\cite{Lopez-Honorez:2013lcm} and average over the redshift dependence to get $f_{dec}$ as a function of the DM mass only, according to Table II of Ref.~\cite{Diamanti:2013bia}. We then modify the energy deposited into the intergalactic medium by inputting the equation above into the  CosmoRec \cite{Chluba:2010ca} package to compute deviations on the ionization history, which now depend on the DM mass and lifetime for a given decay final state. We obtained values for $f_{dec}$ for the $e^+e^-$,$\mu^+\mu^-$,$\tau^+\tau^-$ final states in agreement with Ref.~\cite{Diamanti:2013bia}. In addition, we calculated the efficiencies for decays into quarks and gauge bosons using the recipe of \cite{Slatyer:2012yq}. In detail, in order to compute the bounds for $WW$ and $\bar qq$ final states, we rescale the limits from Ref.~\cite{Lopez-Honorez:2013lcm} by using the appropriate energy deposition efficiency, which in those cases is slightly larger than for the $\mu\mu$ final state, resulting into a slightly more restrictive limit as what shown in  Fig.1, and in agreement with \cite{Slatyer:2012yq}. \\


\begin{figure*}[!t]
\centering
\mbox{\includegraphics[width=\columnwidth]{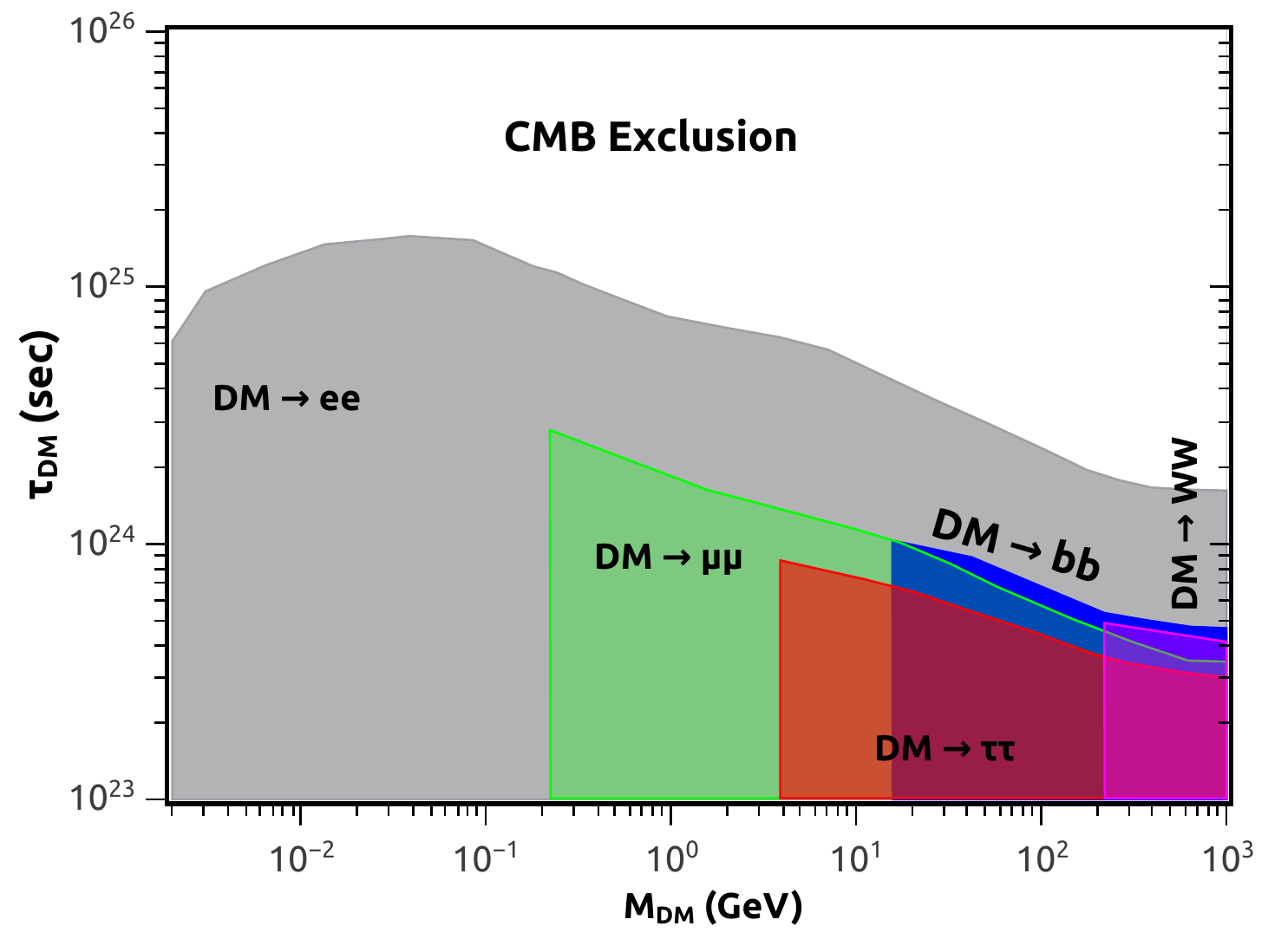}\quad\includegraphics[width=\columnwidth]{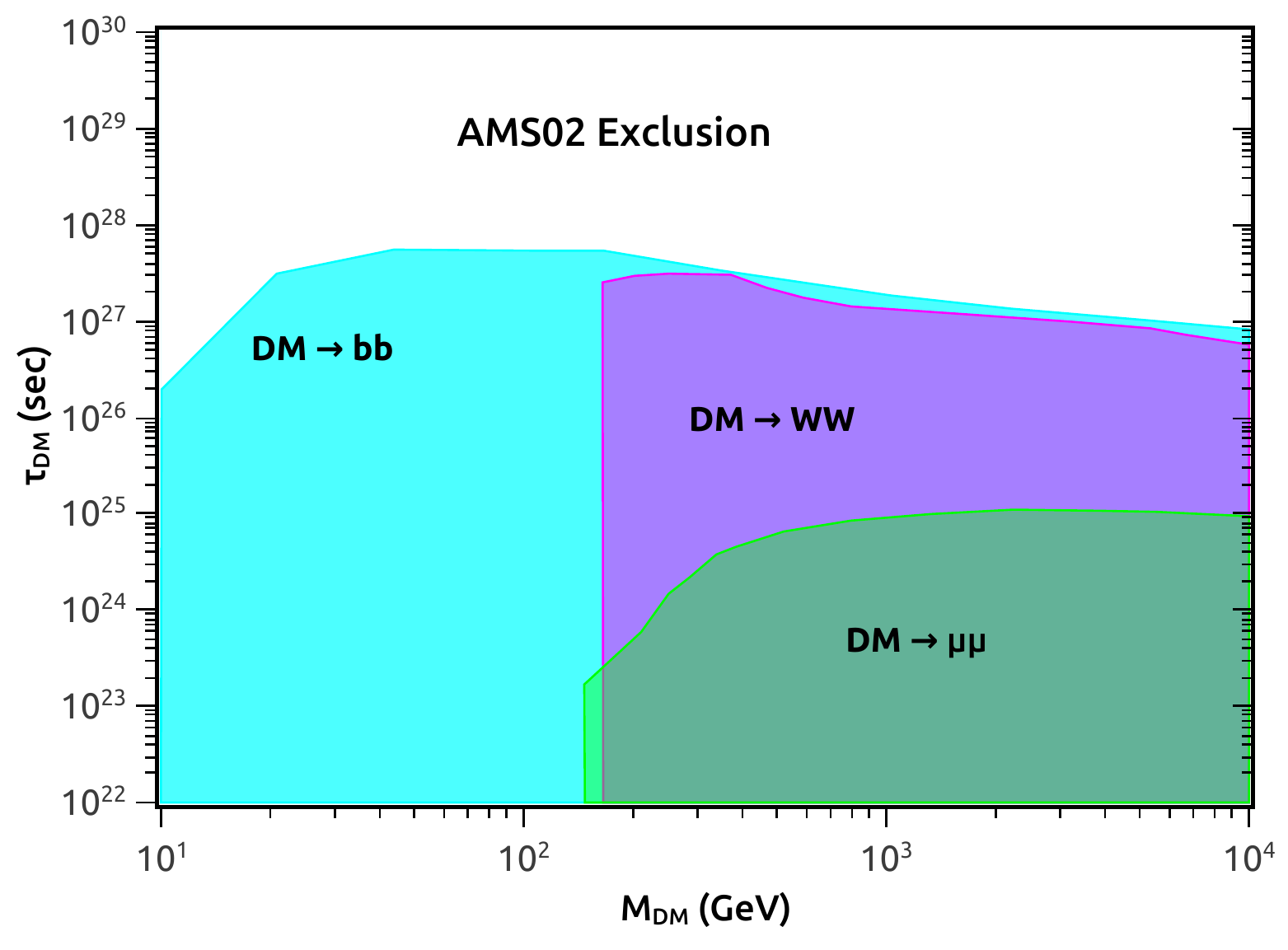}} 
\caption{Model independent bounds on the DM lifetime. {\it Leftmost:} CMB limits for the ee (gray), $\mu\mu$ (green), $\tau\tau$ (orange), bb (blue) and WW (magenta). {\it Left/Center:} Limits from AMS-02 using antiproton data for the bb,WW and $\mu\mu$ channels. {\it Right/Center:} Bounds on the spectral line emission from DM decays. {\it Right:} Constraints on the three-body $ff\nu$ decays mode using measurements of neutrino detectors.}
\label{Graph2}
\end{figure*}

\subsection{Antiproton Data}

Since DM decays produce, in principle, matter and antimatter in equal amount, antiprotons are an interesting target for indirect DM searches, due to the relative rarity of antimatter produced in astrophysical processes. In this section we revisit the procedure to place bounds on the DM lifetime using antiprotons data from AMS-02 \cite{AMS02antiproton}. Antiproton data can be used to set stringent limits on the DM lifetime, since DM decays should at some level produce a sizable amount of antiprotons, even for leptonic final states through the inclusion of electroweak corrections (radiation of a gauge boson which decays hadronically)\cite{Cirelli:2010xx}. \\

With recent AMS-02 precise measurements of the antiproton/proton ($\bar{p}/p$) fraction for energies up to 450 GeV one can derive new restrictive limits on the DM lifetime for several decay modes such as $\bar bb,\,WW$ and $\mu \mu$, since no evidence of new sources of antiprotons were found in the data \footnote{See Ref.~\cite{Hooper:2014ysa} where a claim was put forth about excess antiprotons in the PAMELA data.}. We follow Ref.~\cite{Boudaud:2014qra}, where constraints on the DM annihilation cross section were derived using an older data set for the Einasto DM profile based on the total antiproton flux. Here we will instead obtain limits on the DM lifetime with an NFW profile using the latest AMS data on the $\bar{p}/p$ ratio, and compare our finding with existing limits.\\

The derivation of limits based on antiproton data is subject to large astrophysical uncertainties associated with $\bar{p}$ production, propagation and solar modulation. Here, we employ the standard set of Min-Med-Max propagation models. Min-Med-Max represent values of diffusion parameters which produce a mininum-to-maximum antiproton flux from a DM decay as shown in Table I (See Ref.~\cite{Ibarra:2013cra} for a recent review). Current data seem to disfavor the Min propagation model \cite{Lavalle:2014kca} and the Max-model induces arguably overestimated bounds, so we base our limits on the Med propagation model and an NFW DM distribution.
To obtain limits on the DM lifetime we first solve the  cosmic-ray transport equation in the Galaxy, in a steady state condition for the number density of antiprotons  ($f_{\bar p}$) per unit of kinetic energy $T$,
\begin{equation}
 Q_{\bar p}(T,\vec{r}) + \vec\nabla \cdot [K(T,\vec{r})\vec\nabla f_{\bar p}-
  \vec{V_c}(\vec{r})  f_{\bar p}] -2 h \delta(z) \Gamma_{ann} f_{\bar p}=0,
  \label{transporteq}
\end{equation}where we have neglected energy losses and re-acceleration processes.  We describe below the physical meaning of each of those components:\\
 
{\bf (i)} The first term refers to the primary production of antiproton from DM decays expressed as,

\begin{eqnarray}
Q_{DM} = \frac{\rho}{M_{DM}} \sum_f \Gamma_f \frac{ dN_{f_{\bar p}}}{dK},
\end{eqnarray}where $\rho$ is the DM halo profile assumed to be NFW, and $dN_{f_{\bar p}}/dK$ is the energy spectrum generated using PPPC4DM \cite{Cirelli:2010xx}.

{\bf (ii)}  The second term accounts for the diffusion of cosmic-rays through their propagation in the interstellar medium. It is typically assumed to be constant in the diffusion zone and often parametrized in terms of the particle rigidity (momentum/atomic number) as follows,

\begin{eqnarray}
\mathcal{K}(K)= K_0\, \beta \, ( p /Z)^{\delta},
\end{eqnarray}with $\beta=v/c$, where the normalization ($K_0$) and the spectral index ($\delta$) are associated with the properties of the interstellar medium and derived from measurements of the primary-to-secondary flux ratios of cosmic-rays such as Boron to Carbon \cite{Boudaud:2014qra}, and obviously $Z\equiv 1$ for antiprotons. 

{\bf (iii)} The third term refers to the convection mechanism which accounts for the drift of charged particles away from the disk, assumed to be infinitely thin with a half-height of 100pc \cite{Ibarra:2013cra}, induced by the Galactic Wind with a characteristic velocity $V_{conv}$ and spatially constant in the diffusion zone, i.e., $V_{conv} = sign (z) V_c$. Departures from the thin disk assumption lead to one order of magnitude changes in the final limits as one can see in Fig.6 of Ref.~\cite{Cirelli:2014lwa}.

{\bf (iv)} The fourth term represents the annihilations of antiprotons with the interstellar gas which is proportional to

\begin{equation}
\Gamma_{\bar{p}p} = (n_H + 4^{2/3} n_{He}) \sigma_{\bar{p}p} v_{\bar{p}},
\label{Gammapp}
\end{equation}where, $\sigma_{\bar{p}p} = 0.661 (1+ 0.0115\, T^{-0.774} - 0.948 T^{0.0151} ) b$, for $T< 15.5$~GeV and $\sigma_{\bar{p}p}= 0.036 T^{-0.5} b$ for $T \geq 15.5$~GeV \cite{Tan:1982nc,Protheroe:1981gj}. In Eq.\ref{Gammapp} we assumed that the helium-antiproton annihilation cross section is simply a rescaling of the proton-antiproton \cite{Cirelli:2014lwa}.\\

We now have all ingredients to solve the transport equation and to compute the astrophysical and DM decay predictions for the antiproton flux. A final physical effect, solar modulation, affects the prediction of the antiproton flux at the Earth's atmosphere at energies below $\sim 20$~GeV, as result of the solar cosmic-ray wind and magnetic field.\\

We take into account this effect using the force-field approximation, which determines the antiproton flux as a function of the kinetic energy of the antiproton at the atmosphere ($T_{at}$) by re-scaling the interstellar flux which depends on the antiproton kinetic energy ($T_{is}$) as follows \cite{Gleeson:1968zza}:

\begin{equation}
\Phi_{at} (T_{at}) = \frac{ 2m_p T_{at} + T^2_{at}}{2 m_p T_{is} + T_{is}^2} \Phi^{is} (T_{is}),
\end{equation}with $T_{at}= T_{is}$ - $\phi_F^{\bar{p}}$, where $\phi_F^{\bar{p}}$ is the Fisk potential as given in Table I.\\

Using a data-driven model to account for the proton flux for the energy range of interest as presented by the PAMELA collaboration \cite{Adriani:2011cu}, which is well fitted by a Fisk potential $\phi_F^p=0.7$ (see fifth column of Table 1 of \cite{Cirelli:2014lwa}), one can finally compute the total $\bar{p}/p$ ratio from primary and secondary production processes, as discussed above, { and} enforce the condition that the predicted $\bar p/p$ ratio does not exceed the ratio measured by AMS-02 data \cite{AMS02antiproton} at 95\% C.L, for the specific choices of DM lifetime and mass; That results in constraints on the DM lifetime versus mass plane shown in the second panel of Fig.1. Our results were obtained with PPPC4DMID code \cite{Cirelli:2010xx}.\\

Notice that our limits are competitive with existing ones derived using PAMELA \cite{Ibarra:2013cra} and AMS-02 \cite{Giesen:2015ufa}. In particular, our limits are mildly similar to Ref.~\cite{Giesen:2015ufa} which performed a through analysis by including several energy loss processes we ignored.

\begin{table*}
\begin{center}
Antriproton Propagation Model\\
\begin{tabular}{c c c}
\hline
MED & $\delta=0.7$ & $K_0=0.0112 Kpc^2/Myr$ \\
$V_{c}=12 km/s$ & $L=4$~Kpc & $\phi_F^{\bar{p}}=0.7 GV$\\
\end{tabular}
\caption{Propagation model parameters: $\delta$ and $K_0$ are the spectral index and normalization that go into Eq.6; $V_c$ is the wind velocity in Eq.7; L is the half-height of the cylinder with 20Kpc radius which is used to model the antiproton diffusion. We assumed the proton Fisk potential to be equal to the antiproton which is a good approximation as discussed in Ref.~~\cite{Cirelli:2014lwa}.}
\end{center}
\end{table*}

\begin{figure*}[!t]
\centering
\mbox{\includegraphics[width=\columnwidth]{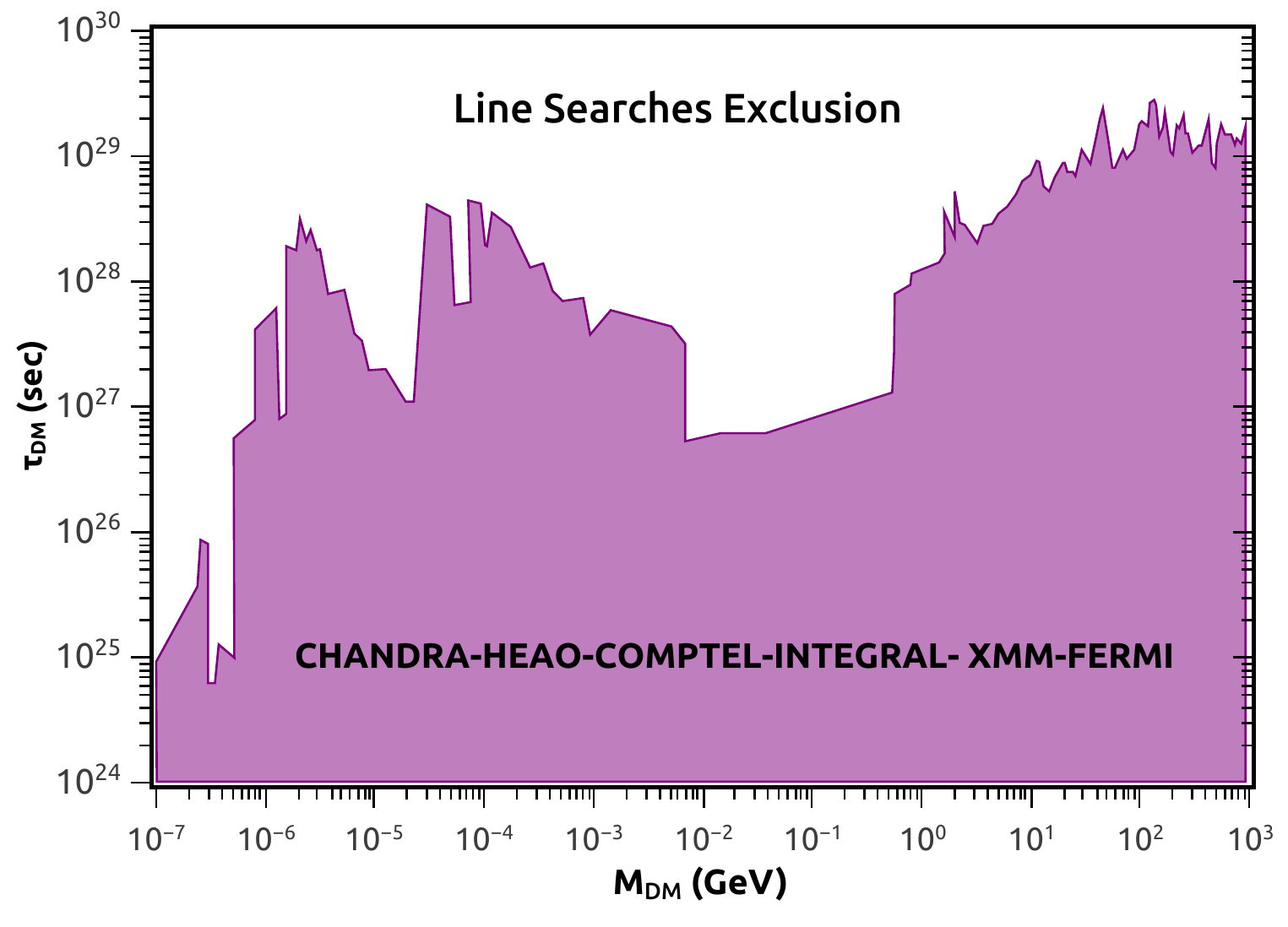}\quad\includegraphics[width=\columnwidth]{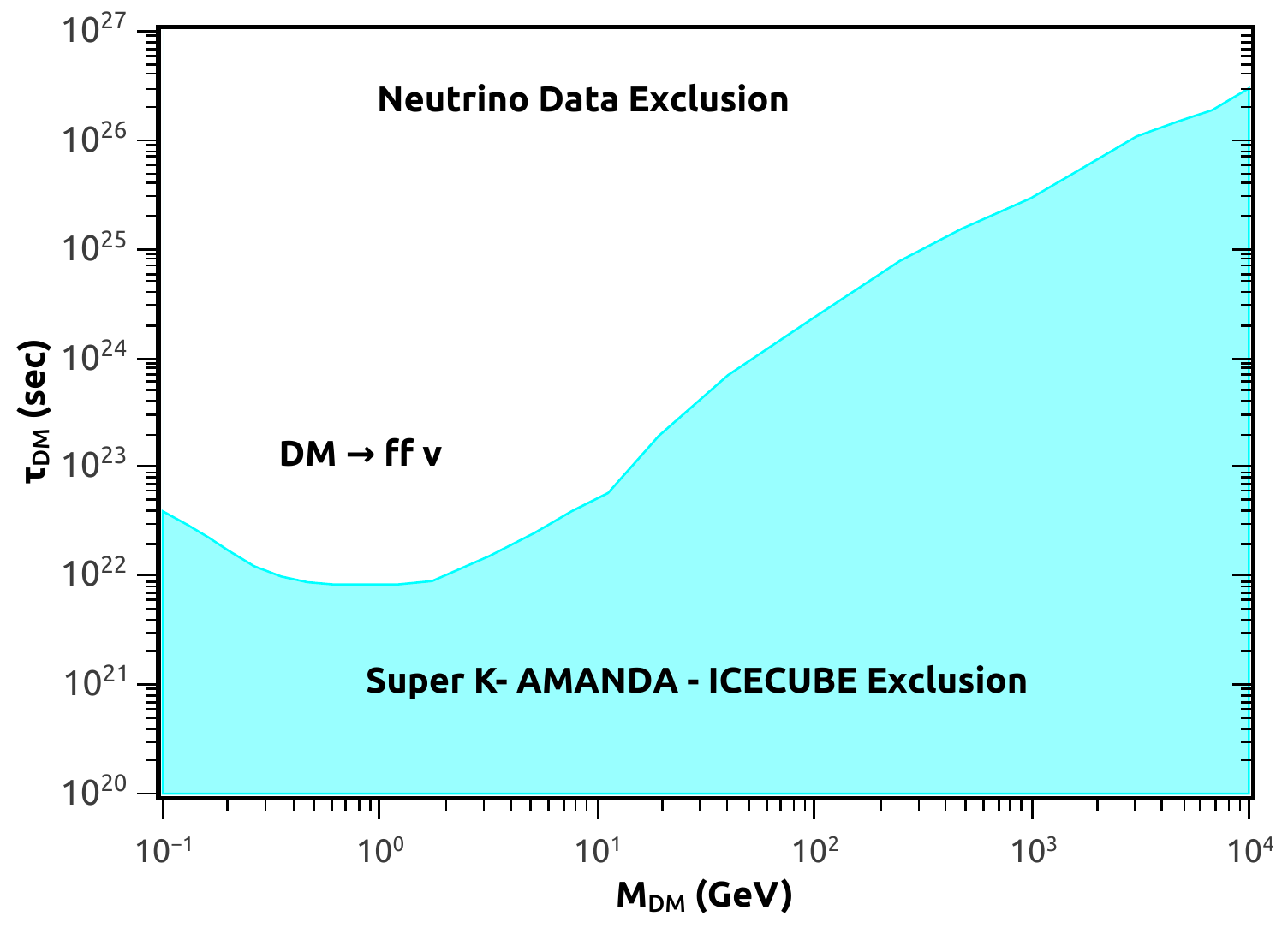}} 
\caption{Model independent bounds on the DM lifetime. {\it Left:} Bounds on the spectral line emission from DM decays. {\it Right:} Constraints on the three-body $ff\nu$ decays mode using measurements of neutrino detectors.}
\label{Graph2_1}
\end{figure*}

\subsection{Line Searches}

If bright enough to be distinguishable from background continuum emission, and if morphologically diffuse \cite{Carlson:2013vka}, gamma-ray spectral lines are known as a veritable smoking gun for DM annihilation or decay signals. Several experiments have searched for line emission at energies between $10^{-7}$GeV up to 400 GeV: (i) Chandra and X-ray Multi-Mirror Mission (XMM) X-ray telescopes cover the 0.007 keV-12 keV range \cite{Boyarsky:2007ay,Bazzocchi:2008fh}; (ii)  High Energy Astronomy Observatory (HEAO) accounts for the 3-48 keV \cite{Boyarsky:2005us}; (iii) INTErnational Gamma-Ray Astrophysics Laboratory (INTEGRAL) the 20k eV-7 MeV \cite{Boyarsky:2007ge}; (iv) The Imaging Compton Telescope (COMPTEL) along with The Energetic Gamma Ray Experiment Telescope (EGRET) screens the MeV-100GeV \cite{Yuksel:2007dr}; (v) Fermi-LAT covering energies up to 462 GeV \cite{Ackermann:2015lka,Ackermann:2012qk}. Here, we simply combine all of those constraints. We point out that we make use here of the latest Fermi-LAT limits on the flux at 95\% C.L for the $180^\circ$ region centered at the Galactic Center (R180), as described in Ref.~\cite{Ackermann:2015lka,Ackermann:2013uma}. Limits on the DM lifetime are obtained after solving for the DM lifetime using the differential flux equation,

\begin{equation}
\tau_{DM} = 16.7 \times 10^{28} {\rm s} \left(\frac{10 GeV}{m_{DM}}\right) \left(\frac{ 10^{-9}cm^{-2}s^{-1}}{\Phi_{\gamma\gamma}}\right)\times J_{decay}.
\end{equation}

The combination of the bounds listed above are shown in Fig.1, third panel. Note that some operators discussed in our work give rise to $Z\gamma$ and $h\gamma$ lines: for those we correct the energy of the gamma-ray line using the relation $E_{\gamma}=M_{DM}(1-m_V/M_{DM})$, where $m_V$ is either the Z or Higgs mass, and divide the lifetime by a factor of two since we have one photon in the final state instead. 

\subsection{Neutrino Data}

Neutrino detectors are sensitive to DM decays and have been used to place limits on the DM lifetime. For a NFW profile for $\rho_{\odot}=0.4$, the full sky differential neutrino flux from DM decays reads \cite{Bertone:2004pz},

\begin{equation}
\frac{d\Phi_{\nu}}{dE_{\nu}} \simeq 1.7 \times 10^{-5} \left( \frac{100 GeV}{ M_{DM}} \right) \left(  \frac{ 10^{24} s}{\tau_{DM}} \right) \frac{dN_{\nu}}{dE_{\nu}} cm^{-2} s^{-1} sr^{-1}.
\end{equation}

Another source of neutrinos from DM decays stems from cosmological decays of DM particle producing a diffuse neutrino flux from decays at all redshifts which reads:

\begin{equation}
\frac{d\Phi_{\nu}}{dE_{\nu}} = \frac{\rho_{DM}}{4\pi M_{DM}} \frac{1}{\tau_{DM}} \int_0^{\infty} dz \left( \frac{1+z}{H(z)}\right) \left( E_{\nu} \frac{dN_{\nu}}{dE_{\nu}} \right) e^{-s_{\nu}(E_{\nu},z)}, 
\end{equation}where $\rho_{DM}$ is the cosmological DM density, $H(z)=H_0\sqrt{\Omega_{\Lambda} + \Omega_{m}(1+z)^3}$ is the expansion rate of the universe, and $s(E_{\nu},z)$ is the universe opacity to neutrinos obtained in Ref.~\cite{AlvarezMuniz:2000es}. The neutrino oscillation probabilities in vacuum is assumed to remain unchanged at the detector. Thus the primary neutrino flux from a specific flavor is redistributed equally into all neutrino flavors, so that the number of expected events is given by,

\begin{equation}
N_{exp}= \left( {\rm time} \times \Delta \Omega \right) \sum_i \int_{E_{min}}^{E_{max}} \frac{d\Phi_{\nu+\bar{\nu}} }{dE} A_{eff}(E_{\nu})dE_{\nu}.
\end{equation}

By comparing with the 95\% C.L limits on the number of events observed, constraints on the DM lifetime for two body decays were derived: (i) Ref.~\cite{Covi:2009xn,PalomaresRuiz:2007ry,Aisati:2015vma} used AMANDA and Super-K data; (ii) Ref.~\cite{Sjostrand:2006za} analyzed recent ICECUBE data; However, operator O14 in Table II induces three body decays ($\bar{f}f\nu$). Hence, we take the limits from those references and use PYTHIA 6.4 \cite{Sjostrand:2006za} to derive the corresponding bounds on three body decay as shown in Fig.\ref{Graph2_1}, rightmost panel. We emphasize that the 95\% C.L. limits were obtained by requiring the theoretical predictions not to overshoot the data in any point.

\subsection{Gamma-ray data}

Observations of the continuous emission of gamma rays give rise to stringent limits on the DM lifetime. Here we employ limits derived from: (i) the extragalactic gamma-ray background, as derived in Ref.~\cite{Ando:2015qda}, which postulates that the sum of the isotropic component from blazars (making up nearly $70\%$ of the total intensity), star-forming galaxies (SFGs), misaligned active galactic nuclei and DM decays  not exceed the measured flux at $95\%$ C.L (Fig.4 of Ref.~\cite{Ando:2015qda}); (ii) limits from Fermi-LAT observations of eight galaxy clusters at gamma-ray frequencies, in $10^{\circ}\times 10^{\circ}$ squared regions centered on the clusters \cite{Huang:2011xr}. We do not duplicate those results here, but we use them in what follows to derive our bounds. For several discussions related to gamma-ray constraints and other topics  which provide complementary limits we point to Refs\cite{Brun:2007tn,Boehm:2009vn,DeLopeAmigo:2009dc,Cholis:2010xb,Garny:2010eg,
Evoli:2011id,Hooper:2012sr,Cirelli:2012ut,Ibarra:2013zia,Jin:2013nta,Gonzalez-Morales:2014eaa,Lovell:2014lea,Cembranos:2014wza,Rott:2014kfa,
Kappl:2015bqa,Chen:2015kla,Jin:2015sqa,Chen:2015cqa,
Hamaguchi:2015wga,Baring:2015sza,Harding:2015bua,Cheng:2015dga}.

\section{Bounds on the Dark Matter Lifetime}

As discussed in the previous section, we assume that continuous global symmetries are broken due to gravitational effects (see Ref.~\cite{Dvali:2015aja} otherwise); in the presence of a continuous global symmetry, one should thus consider Planck-suppressed effective operators which break continuous global symmetries, leading to metastable DM particles. Limits on the lifetime of DM particles from observations in a broad range of frequencies thus allow us to derive general constraints on these operators in settings that advocate continuous global symmetries to stabilize DM candidates.\\

We list in Table II a set of dimension-five  gauge and non-gauge invariant operators that violate continuous global symmetries and induce  DM decay. We point out that our list is not exhaustive or complete; rather, it should serve as a proof of principle, since it includes operators mediating several decay modes that produce significant continuum gamma-ray emission, spectral lines, antiproton, charged leptons and neutrino fluxes; in addition, the set we consider encompasses a variety of DM particle quantum numbers. Here we show results for the gauge-invariant operators only, while in the {\it appendix} we present results for the non-gauge invariant ones. Notice that other functional forms are possible for Planck-scale operators violating continuous global symmetries, for example because of the effects of D-brane instantons, see e.g.Ref.~~ \cite{Blumenhagen:2009qh}.\\

\begin{table}
\begin{center}
 \begin{tabular}{|c|c|}
 \hline
 Name & Interaction Term \\
 \hline
 O1 & $\frac{\lambda_1}{M_{pl}}\bar{f}\gamma^\mu(1+r\gamma_5)f\partial_\mu {\bf S}$ \\
 \hline
O2 & $\frac{\lambda_2}{M_{pl}}{\bf S}F_{\mu\nu}F^{\mu\nu}$\\
 \hline
O3 & $\frac{\lambda_3}{M_{pl}}{\bf  S}\ \epsilon_{\mu\nu\sigma\lambda}F^{\mu\nu}F^{\sigma\lambda}$\\
 \hline
O4 & $\frac{\lambda_4}{M_{pl}}{\bf S}\ G^a_{\mu\nu}G^{a,\mu\nu}$ \\
 \hline
O5 & $\frac{\lambda_5}{M_{pl}}{\bf S}\ \epsilon_{\mu\nu\sigma\lambda}G^{a,\mu\nu}G^{a,\sigma\lambda}$ \\
 \hline
O6 &  $\frac{\lambda_6 m_Z^2}{M_{pl}}{\bf S}\ Z_\mu Z^\mu$\\
 \hline
O7 & $\frac{\lambda_7}{M_{pl}}{\bf S}Z_{\mu\nu}Z^{\mu\nu}$\\
 \hline
O8 & $\frac{\lambda_8}{M_{pl}}{\bf S}\ \epsilon_{\mu\nu\sigma\lambda}Z^{\mu\nu}Z^{\sigma\lambda}$ \\
 \hline
O9 & $\frac{\lambda_9 m_W^2}{M_{pl}}{\bf S}\ W_\mu^+W^{-\mu}$\\
 \hline
O10 & $\frac{\lambda_{10}}{M_{pl}}{\bf S}\ W^+_{\mu\nu}W^{-\mu\nu}$\\
 \hline
O11 & $\frac{\lambda_{11}}{M_{pl}}{\bf S}\ \epsilon_{\mu\nu\sigma\lambda}W^{+\mu\nu}W^{-\sigma\lambda}$ \\
 \hline
O12 & $\frac{\lambda_{12}}{M_{pl}}F^{\mu\nu}Z_\mu\partial_\nu {\bf S}$ \\
 \hline
O13 & $\frac{\lambda_{13}}{M_{pl}}\ \epsilon_{\mu\nu\sigma\lambda}F^{\mu\nu}Z^\sigma\partial^\lambda {\bf S}$\\
 \hline
O14 & $\frac{\lambda_{14}}{M_{pl}} {\bf \bar{\psi}} \tilde{H^{\dagger}} (\slashed{D} L)$\\
 \hline
O15 & $\frac{\lambda_{15}}{M_{pl}} {\bf V}^{\mu} \bar{f} \partial_{\mu} f$\\
 \hline
 O16 & $\frac{\lambda_{16}}{M_{pl}} {\bf V}_{\mu} (H^{\dagger} D_{\nu} H)\, F^{\mu \nu}$\\
  \hline
 \end{tabular}
\caption{Dimension-five Planck-suppressed operators potentially inducing the decay of the DM particle in models with continuous global symmetries. O1-O13 refer to scalar DM (S), O14 to fermion ($\psi$), and O15-O16 to vector (V) DM particles. In the table, $r=\pm 1$, $H$ is the Standard Model Higgs, $\tilde{H}$ the isospin transformation of H, $Z$ and $W$ are the neutral and charged weak gauge bosons, $F_{\mu\nu}$ is the electromagnetic tensor, $G_{\mu\nu}$ is the gluon tensor, {$Z_{\mu \nu}$ and $W_{\mu \nu}$ represent the parts of the SU(2) tensor which involve the physical $Z$ and $W$ gauge bosons respectively}, $f$ and $L$ represent SM fermions. {The list includes both gauge invariant and non-gauge invariant operators} that include a DM particle field singlet under the SM gauge group, but we will restrict our constraints to the gauge invariant ones. See however the appendix for results concerning the non-gauge invariant operators. The dark matter particles which come in multiplets higher dimensional operators are required giving rise to much larger lifetimes avoiding our limits. 
}
\label{spin0decay}
\end{center}
\end{table}

In the Table, we have introduced the dimensionless couplings $\lambda_i\sim{\cal O}(1)$, whose value depends on the unknown mechanism for the quantization of gravity. As we argue below, the precise values of $\lambda_i$ are irrelevant to our conclusions, but we keep the $\lambda_i$'s as free parameters and obtain our limits in the $\lambda$ {\it vs} DM mass plane. For each of the Planck-suppressed operator, we  apply the most stringent limit on the DM lifetime for a given particle mass. Our results are collected in Figures 3-4. In Figure 3 we show the result for gauge invariant operators, whereas in figure 4 for non-gauge invariant one. Notice that several bounds are truncated at some DM mass due to the lack of data at lower energies. Moreover, the sudden change in the behavior of the limits in the figures has to do with the shift in the data set used to constrain a given effect operator. For instance, for operator O1, for masses below $10$~GeV we had to shift from gamma-ray to CMB data that yields much weaker constraints, accounting for the abrupt change in the limit at $M_{DM}=10$~GeV.

Figures 3 shows that models that advocate the presence of continuous global symmetries to stabilize scalar DM candidates might produce a line emission with a very short lifetime (through operator O2/O3), well below the age of the universe, thus ruling out DM masses larger than $100$~keV. It is clear that for any scalar DM operator the whole electroweak WIMP range as well as the large mass range of warm DM is ruled out, since only for $M_{DM} \lesssim 100$~keV are couplings of order one achieved.\\
 
As for fermionic DM candidates stabilized by continuous global symmetries,  operator O14 arises naturally at the Planck scale, yielding an appreciable neutrino and cosmic-ray flux as displayed in the right panel of Fig.3. The result clearly shows that we are able to exclude DM masses above $100$~MeV. Lastly, in models where vector DM particles are stabilized via the existence of continuous global symmetries, operator O15 would automatically be present at the Planck scale, leading to DM decay into fermion pairs. After employing a combination of the bounds shown in Fig.1, we find that masses larger than $10$~MeV induce cosmic-ray and gamma-ray fluxes that exceed the measured values. Conclusions regarding the remaining operators in Table II can be straightforwardly drawn.\\

In summary, we find that dimension-five effective operators at the Planck scale make continuous global symmetries problematic to stabilize DM particles outside very special, restricted mass ranges. In the next section we show how our bounds highly constrain several well-known models in the literature.

\begin{figure*}[!t]
\centering
\mbox{\hspace*{-1cm}\includegraphics[width=\columnwidth]{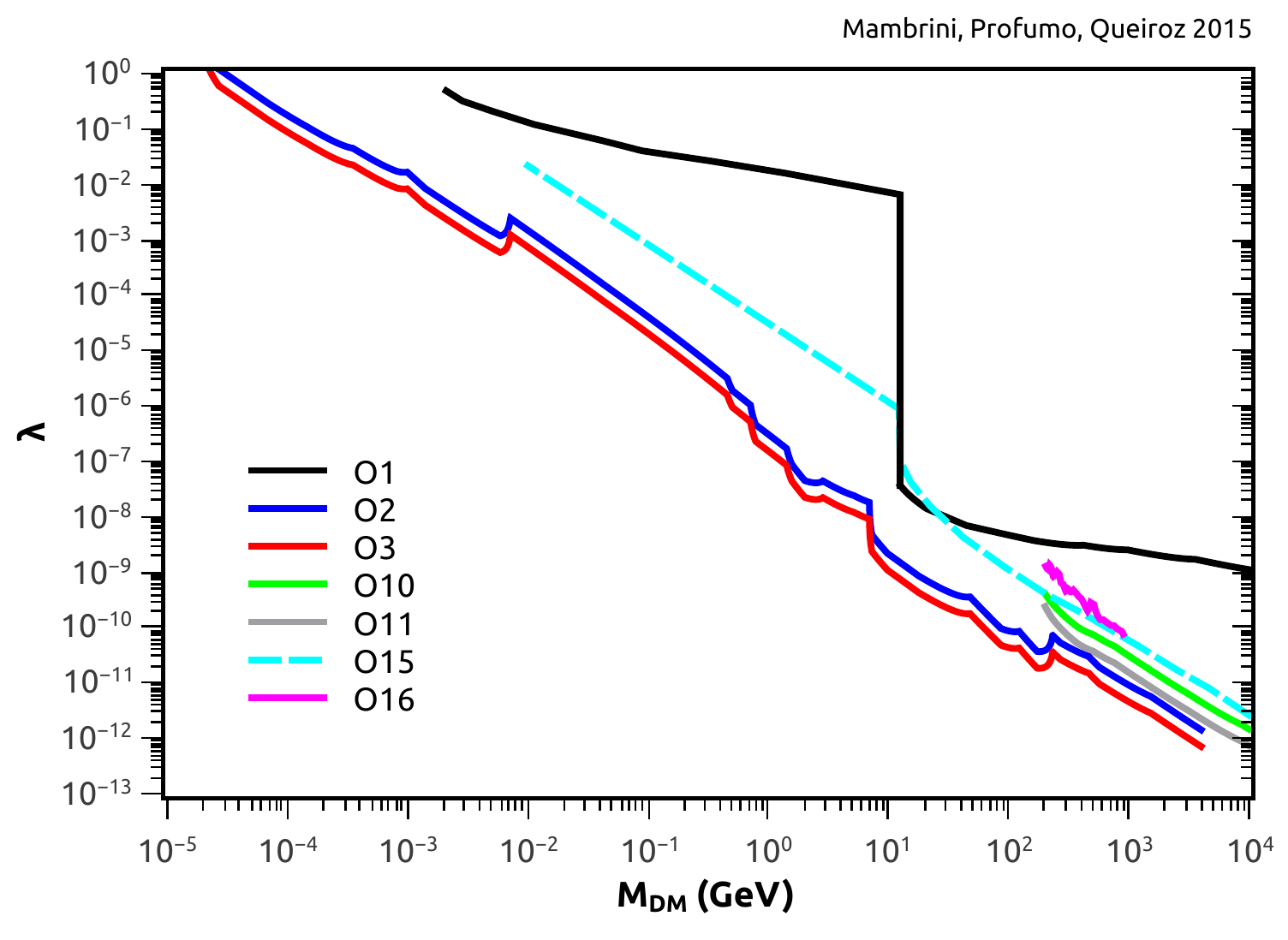}\quad\includegraphics[width=\columnwidth]{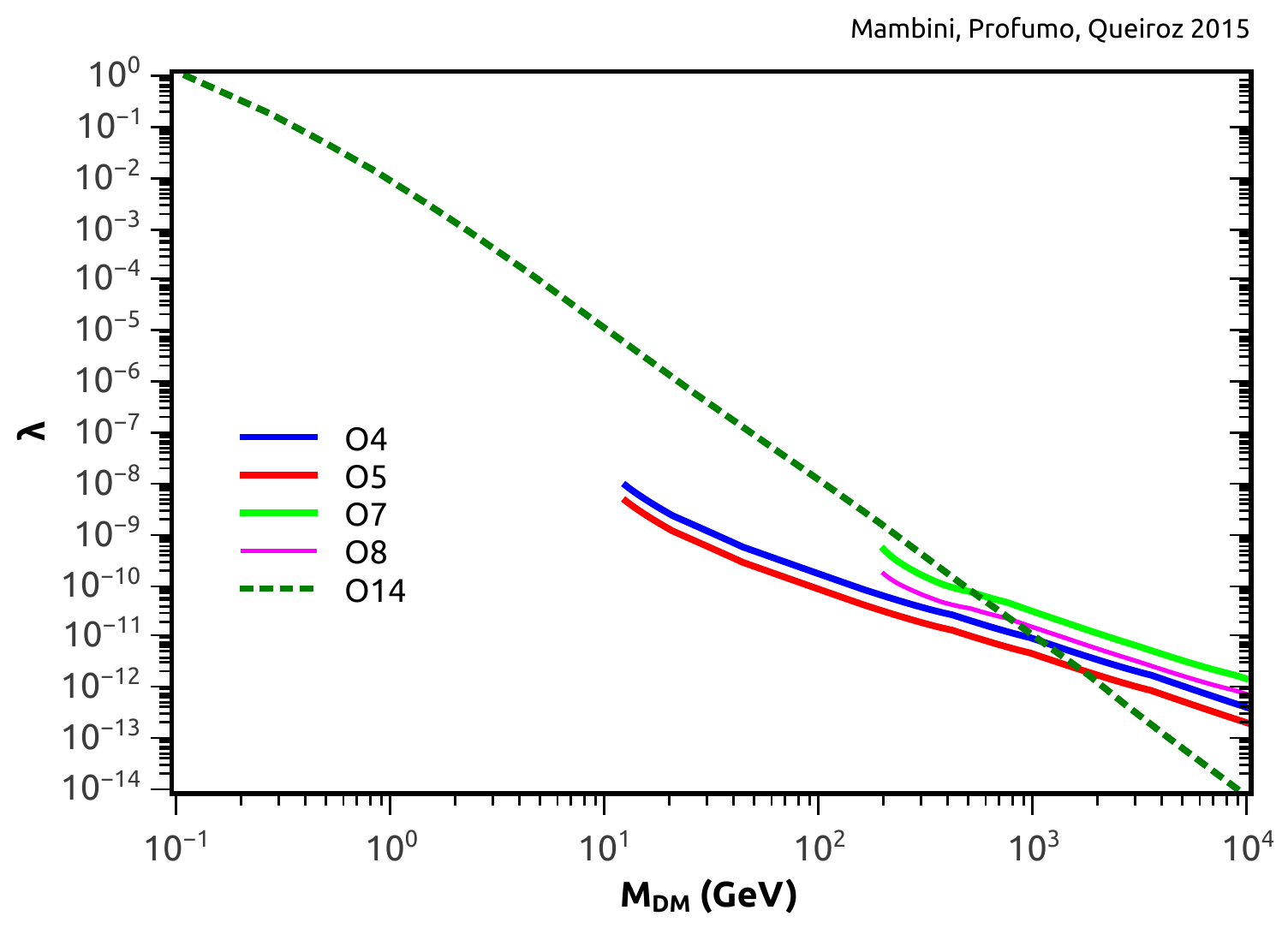}} 
\caption{Limits on the $\lambda$ as a function of the DM plane by enforcing that energy injected at dark ages does not distort the CMB power spectrum at 95\% C.L., and the flux of neutrinos, cosmic-rays and gamma-rays induced by DM decays of each individual operator does not exceed the measured values at $95\%$~C.L. The limits are based on the results presented in Figures 1-2. These bounds are applicable to any scalar, fermion or vector DM particle stabilized by a continuous global symmetry. From our results one may exclude: (i) scalar DM with $M_{DM} > 100$~keV; (ii) fermion DM with $M_{DM} > 100$~MeV; (iii) Vector DM with $M_{DM} > 10$~MeV.}
\label{Graph3}
\end{figure*}

\section{Concrete Models}
In what follows, we work under four assumptions: (i) global symmetries are in fact broken at the Planck scale;(ii)  O(1) non-renormalizable couplings are a good approximation; (iii) dark matter field is a singlet field; (iv) the dark matter density distribution is well described by a NFW profile. 

It is important to keep in mind as caveats that despite the existence of plausible arguments ruling out global symmetries at the Planck Scale, the validity of such arguments is still under debate. Moreover, in general continuous global symmetries can be replaced by other symmetries such as discrete symmetries, circumventing our constraints. We assume $O(1)$ non-renormalizable couplings for the non-renormalizable operators mediating decay. Any departure from this assumption is easily factoring in by rescaling using the expressions for the width rate given in Table 3.

We apply here the results from the previous sections to concrete models, widely discussed in the literature where the dark matter field is a singlet field, for which our bounds are directly applicable.

\begin{center}
{\it Left-Right Model}
\end{center}

We consider the left-right mirror symmetric model with the continuous global symmetry $U(1)_{B-L}\otimes U(1)_X$ of Ref.~~\cite{Yang:2013bda,Yang:2014jca}. There, the continuous global symmetry prohibits the term $\bar{L} H \psi_R$, where $\psi$ is a fermionic DM candidate. However, such symmetry is generically violated at the Planck scale and therefore the O14 operator ought to exist, thus ruling out DM masses above 100 MeV. Ref.~~\cite{Yang:2013bda,Yang:2014jca} also invokes the case of WIMP scalar DM protected by the continuous global symmetry, but once again, as we see in the left panel of Fig.2, the entire corresponding WIMP mass range is excluded. As a result, the model described in Ref.~\cite{Yang:2013bda,Yang:2014jca} does not appear to have a plausible DM candidate. Unless the invoked continuous global symmetry can be replaced by a $Z_N$ discrete symmetry of some sort, the model is strongly disfavored by data.

\begin{center}
{\it Two Higgs Doublet Model}\\
\end{center}

In the original two Higgs doublet model no DM candidate is present. Nevertheless, if the second Higgs doublet is odd under a $Z_2$ symmetry the CP-even scalar of that doublet can be a DM candidate. This is the case in the so-called Inert Two Higgs Doublet Model (I2HDM) \cite{LopezHonorez:2006gr}. Recently, a continuous global symmetry has been proposed to replace the $Z_2$ symmetry \cite{Bhattacharyya:2013rya}. The authors focus on the spontaneous symmetry breaking of the continuous global symmetry through the {\it vev} of the second Higgs doublet and comment on the possibility of having an unbroken continuous global symmetry. { Despite the interesting Higgs physics implications produced by the use of the continuous global symmetry \cite{Bhattacharyya:2013rya}, operator O2 should be present at the Planck scale; thus, from Fig. 2 we conclude that DM masses larger than $100$~keV are problematic along with the possibility of having viable WIMP DM candidates in the model.}

\begin{center}
{\it Singlet Fermion Model}\\
\end{center}

The minimal fermionic DM model studied in Ref.~\cite{Baek:2013ywa} advocates a continuous global symmetry responsible for stabilizing a singlet fermion which yields the desired thermal relic abundance and is consistent with direct searches. The Planck suppressed effective operator O14, however, rules out the entire WIMP mass range. As we mentioned before, in principle one could replace the continuous global symmetry by a discrete symmetry, since $Z_N$ is a subgroup of $U(1)$. However, the necessary discrete symmetry might turn out to imply a rather large and unnatural tuning of the model.

\begin{center}
{\it Radiative See-Saw Model}\\
\end{center}

A radiative lepton model in which the charged lepton masses are
generated at one-loop level whereas and the neutrino masses at two-loop level has been proposed in Ref.~~\cite{Baek:2014awa}. In this model the continuous global and $Z_2$ symmetry have been invoked and two DM candidates postulated. A singlet fermion, referred to as $n^{\prime}$ in Table I of Ref.~\cite{Baek:2014awa}, is not odd under the $Z_2$ symmetry, and claimed to be a WIMP due to the presence of a continuous global symmetry. Similarly to the previous model, Planck-suppressed dimension-five operators {\bf discard} such possibility. 

\begin{center}
{\it Zee-Babu Model}\\
\end{center}

The Zee-Babu model adds to the SM a singly-charged and a doubly-charged scalar \cite{Lindner:2011it}. Recently, an extension of the Zee-Babu model has been put forth by adding  a singlet fermion which is stabilized by a continuous global $U(1)_{B-L}$ symmetry. This continuous global symmetry also forbids terms like $\bar{L} \tilde{H} N$. There, the neutral fermion does not carry a lepton number so it is purely a neutral fermion. Nevertheless, as we discussed, this continuous global symmetry does not hold up to the Planck scale and consequently the operator O14 arises, inducing an excess production of neutrinos, gamma-ray and comic-rays, which results into the exclusion of DM masses below $100$~MeV, in tension with what presented in Figs.1-2 of Ref.~\citep{Lindner:2011it}.

\begin{center}
{\it 3-3-1 Models}\\
\end{center}

3-3-1 models refer to $S(3)_c \otimes S(3)_L \otimes U(1)_N$ gauge extensions of the Standard Model \cite{Foot:1992rh}. In Ref.~\cite{Mizukoshi:2010ky} a continuous global symmetry with the purpose of avoiding undesirable mixing among the gauge bosons and of guaranteeing that the lightest particle charged under the continuous global symmetry be stable. Both a complex scalar and a heavy Dirac fermion were studied as potential  WIMP DM candidates. In a similar vein to what discussed above, the WIMP mass regime in this model is in jeopardy due to the aforementioned gravity effects. In the urge of preventing the use of continuous global symmetries in the model, Ref.~\cite{Dong:2014wsa} proposed adding an extra gauge symmetry, which would completely change the associated DM phenomenology.

 \section{Conclusions}

Based on general lessons from quantum mechanics and general relativity, it reasonable to assume that no global symmetries are allowed in a consistent theory of quantum gravity. We have presnted, in a model-independent approach, robust gamma-, X-ray, CMB, and cosmic-ray constraints on decaying DM particles, using a large set of data, including data from Fermi-LAT, AMS-02, Super-Kamiokande, Planck, WMAP9, AMANDA, and Icecube among others. We have then applied those bounds to scalar, vector and fermion DM particles decaying through dimension-five Planck-suppressed effective operators. We stress that our findings are based on the following assumptions: 

(i) global symmetries are broken at the Planck scale;

(ii)  the operators mediating dark matter decay have $O(1)$ non-renormalizable couplings; 

(iii) the dark matter particle is represented by a singlet field; 

(iv) the dark matter density distribution in the Galaxy is well described by a NFW profile.\\ 

Under these assumptions, we have derived the following constraints on the possible mass range:

(i) scalar DM: $M_{DM}  \lesssim 100$~keV  ;

(ii) fermionic DM: $M_{DM} \lesssim 100$~MeV ;

(iii) Vector DM: $M_{DM} \lesssim 10$~MeV . \\

Lastly, we have applied our limits to instances of parameter space of Left-Right, Two-Higgs Doublet, Singlet Fermionic, Zee-Babu, 3-3-1 and Radiative See-Saw models to conclude that the occurrence of DM particles in such models is generically problematic outside the DM particle mass ranges listed above. Our results basically rule out the entire WIMP mass range for models invoking continuous global symmetries to stabilize the DM particles.

We emphasize that our results  rely on the critical assumptions listed above, especially on the hypothesis that continuous global symmetries are violated at the Planck scale. Notice that if one uses non-renormalizable couplings smaller than unit, our limits can be simply rescaled using Table 3 containing the decay rates. Moreover, if one adopted a different density profile that would quantitatively change the acceptable DM mass ranges, but leave the overall conclusions unchanged. However, if the DM particle belongs to a multiplet, and thus only higher dimension operators induce the dark matter decay, our limits would not be  relevant. In summary, our bounds and conclusions can be subject to significant changes.

\section*{Acknowledgement}

The authors thank Michel Dine, Dan Hooper, Herbi Dreiner, Pavel Fileviez Perez, Alejandro Ibarra, Paul Langacker, Will Shepherd, Christoph Weniger, and, especially, Marco Cirelli for useful discussions. We are indebted to Tom Banks for important comments and reading the manuscript in early stages. Y.M was partially supported by the Spanish MICINN's Consolider-Ingenio 2010 Programme under grant
Multi-Dark { CSD2009-00064}, the contract { FPA2010-17747}; European Union FP7 ITN INVISIBLES (Marie Curie Actions, { PITN-GA-2011- 289442}); ERC advanced grants Higgs\@LHC and MassTeV; Research Executive Agency (REA) of the European Union under the Grant Agreement
{ PITN-GA2012-316704} ("HiggsTools"); and LIA-TCAP of CNRS. SP is partly supported by
US Department of Energy Award SC0010107. FSQ thanks UCSC for its hospitality and funding from US Department of Energy Award SC0010107 during early stages of this work. 

\section{Appendix:Decay widths and non-gauge invariant operators}

\begin{figure}[!t]
\centering
\includegraphics[width=\columnwidth]{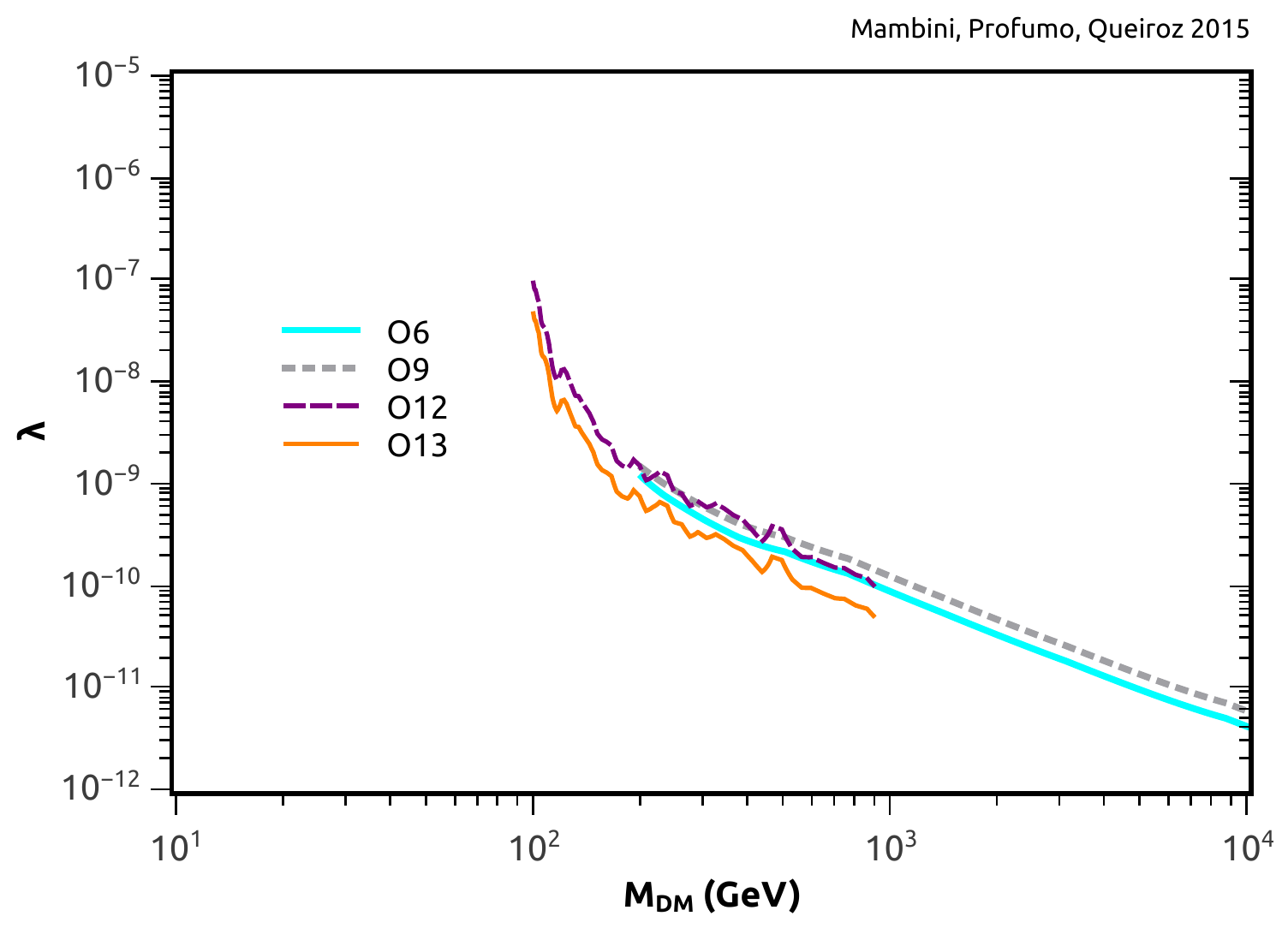} 
\caption{Limits from non-gauge invariant operators on $\lambda$ as a function of the DM mass, obtained by enforcing the $95\%$~C.L. bounds from the various observational probes discussed above.}
\label{Graph10}
\end{figure}

 In Table III we list the decay width associated with each operator discussed in the manuscript; in addition we discuss some non-gauge invariant operators (O6-O9-O12-O13) that might appear in more complex setups such as non-Abelian theories \cite{Gross:2015cwa}. We also show the limits stemming from such operators in Fig.\ref{Graph10}. 

\begin{table*}[t]
\begin{center}
\scalebox{1}{
\begin{tabular}{|c|c|c|}
 \hline
 Name & Interaction term &  Decay Rate \\
 \hline
 O1 & $\frac{\lambda_1}{M_{pl}}\bar{f}\gamma^\mu(1+r \gamma_5)f\partial_\mu {\bf S}$&  $\Gamma(S \to f \bar{f})$ =$\frac{\lambda^2\,r^2 N_f}{2\pi}\frac{m_f^2m_S}{M_{pl}^2}A^{1/2}$ \\
 \hline
O2 & $\frac{\lambda_2}{M_{pl}}{\bf S}F_{\mu\nu}F^{\mu\nu}$ & $\Gamma(S \to \gamma\gamma)$ = $\frac{\lambda^2m_S^3}{4\pi M_{pl}^2}$ \\
 \hline
O3 & $\frac{\lambda_3}{M_{pl}}{\bf  S}\ \epsilon_{\mu\nu\sigma\lambda}F^{\mu\nu}F^{\sigma\lambda}$& $\Gamma(S \to \gamma\gamma)$= $\frac{\lambda^2}{\pi M_{pl}^2}m_S^3$ \\
 \hline
O4 & $\frac{\lambda_4}{M_{pl}}{\bf S}\ G^a_{\mu\nu}G^{a,\mu\nu}$ & $\Gamma(S \to gg)$ =$\frac{\lambda^2m_S^3 N_g}{4\pi M_{pl}^2}$ \\
 \hline
O5 & $\frac{\lambda_5}{M_{pl}}{\bf S}\ \epsilon_{\mu\nu\sigma\lambda}G^{a,\mu\nu}G^{a,\sigma\lambda}$& $\Gamma(S \to gg)$=$\frac{\lambda^2}{\pi M_{pl}^2}N_g m_S^3$ \\
 \hline
O6 & $\frac{\lambda_6 m_Z^2}{M_{pl}}{\bf S}\ Z_\mu Z^\mu$&  $\Gamma(S \to ZZ)$=$\frac{\lambda^2m_S^3}{32\pi M_{pl}^2}A^{1/2}\times \left (A+12\frac{m_Z^4}{m_S^4}\right )$ \\
 \hline
O7 & $\frac{\lambda_7}{M_{pl}}{\bf S}Z_{\mu\nu}Z^{\mu\nu}$& $\Gamma(S \to ZZ)$=$\frac{\lambda^2m_S^3}{4\pi M_{pl}^2}A^{1/2}\times \left (A+6\frac{m_Z^4}{m_S^4}\right )$ \\
 \hline
O8 &$\frac{\lambda_8}{M_{pl}}{\bf S}\ \epsilon_{\mu\nu\sigma\lambda}Z^{\mu\nu}Z^{\sigma\lambda}$ &  $\Gamma(S \to ZZ)$=$\frac{\lambda^2m_S^3}{\pi M_{pl}^2}A^{3/2}$ \\
 \hline
O9 & $\frac{\lambda_9 m_W^2}{M_{pl}}{\bf S}\ W_\mu^+W^{-\mu}$& $\Gamma(S \to W^+W^-)$=$\frac{\lambda^2m_S}{64\pi M_{pl}^2}A^{1/2}\times \left (A+12\frac{m_W^4}{m_S^4}\right )$ \\
 \hline
O10 & $\frac{\lambda_{10}}{M_{pl}}{\bf S}\ W^+_{\mu\nu}W^{-\mu\nu}$&  $\Gamma(S \to W^+W^-)$=$ \frac{\lambda^2 m_S^3}{4\pi M_{pl}^2}A^{1/2}\times \left (A+6\frac{m_W^4}{m_S^4}\right )$ \\
 \hline
O11 & $\frac{\lambda_{11}}{M_{pl}}{\bf S}\ \epsilon_{\mu\nu\sigma\lambda}W^{+\mu\nu}W^{-\sigma\lambda}$ & $\Gamma(S \to W^+W^-)$= $\frac{g_{0p}^2m_S^3}{\pi M_{pl}^2}A^{3/2}$ \\
 \hline
O12 & $\frac{\lambda_{12}}{M_{pl}}F^{\mu\nu}Z_\mu\partial_\nu {\bf S}$ & $\Gamma(S \to Z\gamma)$=$\frac{\lambda^2m_S^3}{32\pi M_{pl}^2}\left (1-\frac{m_Z^2}{m_S^2}\right )^3$ \\
 \hline
O13 & $\frac{\lambda_{13}}{M_{pl}}\ \epsilon_{\mu\nu\sigma\lambda}F^{\mu\nu}Z^\sigma\partial^\lambda {\bf S}$& $\Gamma(S \to Z\gamma)$=$\frac{\lambda^2m_S^3}{8\pi M_{pl}^2}\left (1-\frac{m_Z^2}{m_S^2}\right )^3$ \\
 \hline
O14 & $\frac{\lambda_{14}}{M_{pl}} {\bf \bar{\psi}} \tilde{H^{\dagger}} (\slashed{D} L)$ & $\Gamma(\psi \to ff \nu)$=$\frac{\lambda^2 v^2 G_f^2 M_{\psi}^5}{192 \pi^3 M_{pl}^2}$\\
 \hline
O15 & $\frac{\lambda_{15}}{M_{pl}} {\bf V}^{\mu} \bar{f} \partial_{\mu} f$ & $\Gamma(V \to ff)$=$\frac{\lambda^2 M_V^3}{ 64 \pi M_{pl}^2}A^{3/2}$ \\
 \hline
 O16 & $\frac{\lambda_{16}}{M_{pl}} {\bf V}_{\mu} (H^{\dagger} D_{\nu} H)\, F^{\mu \nu}$ & $\Gamma(V \to H \gamma)$=$\frac{\lambda^2 v^2 M_V^3}{ 64 \pi M_{pl}^2 m_h^2}\left (1-\frac{m_h^2}{m_V^2}\right )^3$ \\
  \hline
 \end{tabular}}
\caption{Decay width of the dimension five planck suppressed operators, where $r=\pm 1$, $A=1-4 m^2/M_{DM}^2$, with m being the mass of the final state particle, $G_f$ the fermi constant, $m_h$ higgs mass, and $v=246$~GeV.}\label{spin0decay}
\end{center}
\end{table*}

\bibliographystyle{apsrev4-1}
\bibliography{darkmatter}

\end{document}